# Atomic structure analysis of nanocrystalline Boehmite AlO(OH)

Stefan Brühne[*a,b], Saskia Gottlieb[a], Wolf Assmus[a], Edith Alig[b], Martin U. Schmidt[b]

[a] Physikalisches Institut, Johann Wolfgang Goethe-Universität, Max-von-Laue-Str. 1, D-60438 Frankfurt am Main, Germany
[b] Institut für Anorganische und Analytische Chemie, Johann Wolfgang Goethe-Universität, Max-von-Laue-Str. 7, D-60438 Frankfurt am Main, Germany

* corresponding author
*bruehne@physik.uni-frankfurt.de*

**Abstract**

Nanocrystalline *n*-AlO(OH) (Disperal P2® of Sasol) was investigated by means of the atomic pair distribution function (PDF). The PDF is derived from powder diffraction data, an ideally resolved PDF is obtained from a synchrotron source which provides a large maximal scattering vector length $Q_{max} > 300$ nm$^{-1}$. Here, however, we were able to reveal atomic structure details of the ~ 4 nm particles from in-house diffraction data ($Q_{max} = 130$ nm$^{-1}$): PDF least squares model refinements show that *n*-AlO(OH) is of the layered Boehmite structure type (oC16, *Cmcm*). But the structure is uniformly distorted in domains of ~ 2 nm size within the nano particles. The hydrogen bonds between the layers are widened up significantly by +13 pm, accounting for the higher reactivity when compared to microcrystalline Boehmite. Our results from only one "nanocrystallographic" experiment are consistent with a trend which was found *via* the Rietveld technique on a series of AlO(OH) of different crystallite size (Bokhimi *et al.*, 2001 [2]). In addition, the PDF contains information on structural distortion as a function of (coherence) domain size within the nano particles.

**Introduction**

Nanocrystalline metal oxides like nano-(*n*-)ZnO, *n*-ZrO$_2$, *n*-MgO, *n*-Al$_2$O$_3$ or *n*-Fe$_3$O$_4$ have been of growing interest in the recent decade due to their unique properties differing from those of the respective microcrystalline forms. A number of them is commercially available, *e.g.* a series of grades of pure nanocrystalline Boehmite, *n*-AlO(OH), is distributed by Sasol under the trade name Disperal®. The crystallites reach sizes as small as 4.5 nm [1]. Due to the high dispersability of *n*-AlO(OH), it is used in many applications such as to produce sol-gel Al$_2$O$_3$-ceramics, Al$_2$O$_3$-supported catalysts, refractory materials, in rheology control or in surface frictioning and paint detackification.

There is, by definition, no way to analyze single crystal X-ray diffraction data from such nano-materials and the accurate atomic structure thereby remains obscured. Even in X-ray powder diffractograms the confined crystallite size excessively broadens the peaks. Therefore structure information from Rietveld refinement is difficult to be obtained since this method assumes long-range structural coherence in the material. Nevertheless, there have been attempts to extract as much information as possible from a series of *n*-AlO(OH) grades via the Rietveld technique [2].

As an alternative, the atomic pair distribution function (PDF) simultaneously contains information on different length scales as it uses the total (*i.e.* both Bragg and diffuse) powder scattering intensity [3]. It accounts for the real local atomic structure since it does not base on long-range periodicity. A prominent example for a nano-crystal investigated by PDF is *e.g.* ZnS [4]. For the Al−O(−H) system, recently a γ-$Al_2O_3$ ceramic produced *via* a sol-gel process was analyzed by means of PDF refinement from synchrotron data. Interestingly, on a length scale of ~1 nm, features of the precursor material AlO(OH) have been retained in the bulk γ-$Al_2O_3$ [5].

In our contribution we present the PDF analysis as a straightforward method to get detailed insight into the atomic structure of nano-materials such as *n*-AlO(OH). In this case the use of in-house X-ray Guinier powder diffraction data was possible – though being of lower resolution than synchrotron data, they are obtained in a comfortable way.

**Experimental**

For the investigations commercially available nano-crystalline Boehmite powder, AlO(OH) of high purity (0.005-0.015% $Fe_2O_3$), Disperal® P2, [1] was used. The crystallite size is given as 4.5 nm and the particle size as measured on the powder by laser diffraction is $d_{50}$ = 45 μm [1]. Scanning electron microscopy was applied on a Au-vaporised specimen to verify spherical aggregates of size from 2 to 30 μm (Fig. 1). In Fig. 1 one also can detect spherical features of less than 0.5 μm in diameter inside the aggregates.

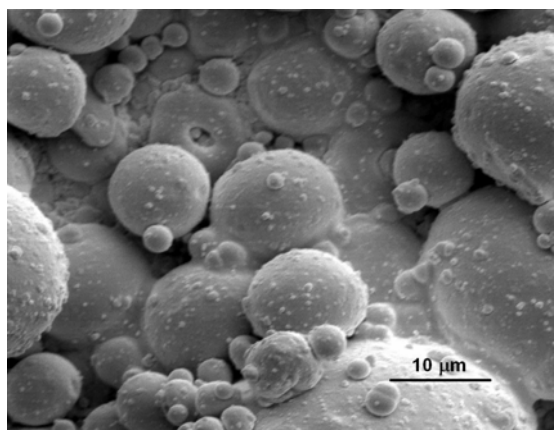

Figure 1: Scanning electron micrographs showing spherical aggregates of *d* = 2 to 30 μm.

X-ray powder diffraction was performed in transmission geometry on STOE Stadi P using Ge(111) monochromatized Cu K$\alpha_1$ radiation and a linear PSD. The diffractogram is shown in Fig. 2a, typical are the broad refections on a high background. The pattern can be indexed with the *C*-centred orthorhombic unit cell of Boehmite mineral (*Cmcm*, *a* = 0.28679(3) nm, *b* = 1.214(1) nm, *c* = 0.36936(3) nm [6]). The full widths at half maximum (FWHM) of the reflections range from 3.6 to 5.4°/2θ. Using the Scherrer equation (1) with *k* = 1 and $\lambda = 0.154$ nm, an average crystallite size can be determined to $D_{\text{Scherrer}} \sim 4.3(8)$ nm (Tab. 1) which is consistent with [1].

$$D_{\text{Scherrer}} = k \cdot \lambda \cdot 57.3 \, / \, [\text{FWHM} \cdot \cos(\theta)] \quad (1)$$

Another measurement using Si(111) monochromatized Mo K$\alpha_1$ radiation was obtained by averaging of 5 scans ($2\theta = 4$ to $100°$, 13 h each) from a Huber-Guinier diffractometer (flat plate transmission) using a scintillation point detector. Special care has been taken to obtain good counting statistics at high $2\theta$ for the subsequent PDF calculations. The background corrected result is plotted in Fig. 2b – still, at high angles, broad but strongly overlapping reflections are visible. Due to their strong overlap they are not uniquely resolvable in a Rietveld analysis as they are corrupted by the diffuse background arising from the sample over the whole $2\theta$ range. However, the atomic pair distribution function (PDF) contains information from these features.

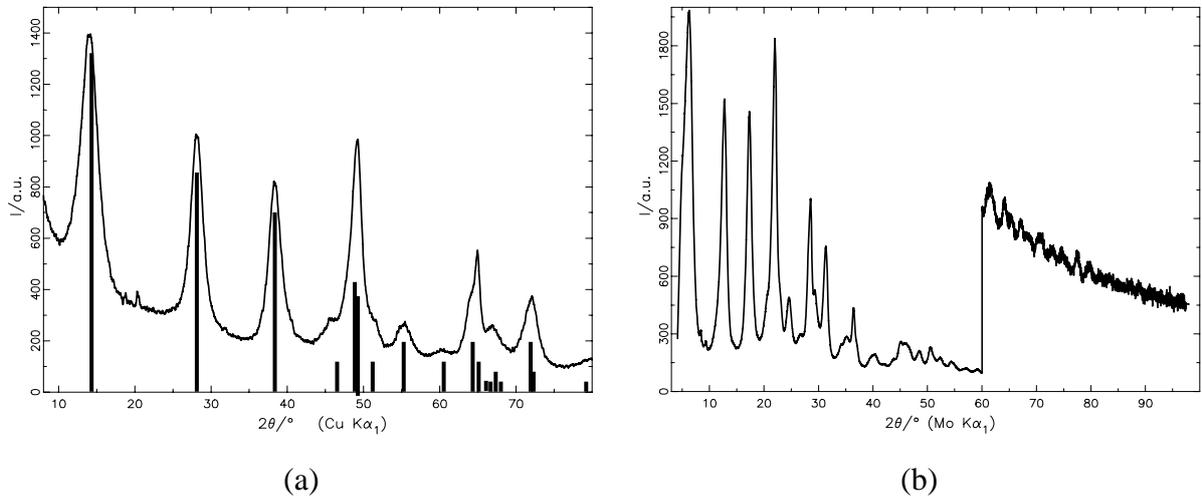

(a)                                                (b)

Figure 2: X-ray powder diffractograms of *n*-AlO(OH), (a) Cu K$\alpha_1$ radiation, bars indicate Boehmite reflections consistent with [6]; (b) Mo K$\alpha_1$ radiation (for $2\theta > 60°$ the intensity scale is enhanced by factor 10 in (b))

Table 1: FWHM analysis of the Cu K$\alpha_1$ data in Fig. 2a using equation (1) ($k = 1$ and $\lambda = 0.154$ nm) to estimate the crystallite size

| $h\,k\,l$ | peak position $2\theta/°$ | $\theta/°$ | FWHM/° | $D_{\text{Scherrer}}$/nm |
|---|---|---|---|---|
| 0 2 0 | 14.1 | 7.1 | 2.67 | 3.33 |
| 0 2 1 | 28.4 | 14.2 | 1.88 | 4.81 |
| 1 3 0 | 38.4 | 19.2 | 1.97 | 4.74 |
| average ($D \pm$ standard deviation)/nm |  |  |  | $4.3 \pm 0.8$ |

The PDF gives the probability to find any atom i (*i.e.* in the X-ray case a maximum of the electron density $\rho$) in the distance $r_{ij}$ to any other atom j normalised to the average electron density $\rho_0$ in a structure (2).

$$G(r) = 4\pi r \, [\rho(r) - \rho_0] \quad (2)$$

The experimental PDF, $G_{\text{exp}}(r)$ is the direct sine Fourier transform (3) of the normalized scattering intensity $S(Q)$. The scattering vector length is $Q = |\mathbf{Q}| = 4\pi \sin(\theta)/\lambda$ where $\theta$ is the scattering angle and $\lambda$ the wavelength of the used radiation. $S(Q)$ is obtained from the total scattering intensity $I(Q)$ by experimental background subtraction, Laue and polarisation correction and subsequent normalisation w. r. t. an average scattering factor of the material under investigation [3].

$$G_{exp}(r) \;=\; 2/\pi \int_0^\infty Q[S(Q) - 1]\sin(Q \cdot r)\,dQ \qquad (3)$$

The real-space reolution in $r$ of a PDF strongly depends on the maximal scattering vector length $Q_{max}$. Using Mo K$\alpha_1$ radiation and $2\theta_{max} = 100°$ yields $Q_{max} \approx 130$ nm$^{-1}$. Good instrumental resolution in $Q$ minimizes the exponential fall-off of the PDF with $r$. For data acquistion from the Mo K$\alpha_1$ data $I(Q)$, PDFgetX2 software [5] was used.

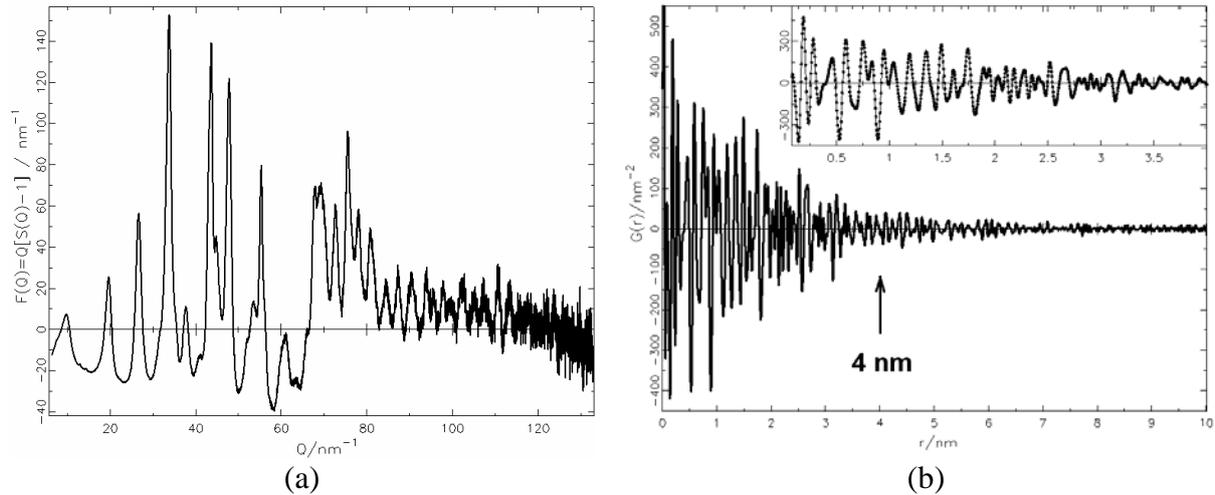

(a)                          (b)

Figure 3: (a) structure function of $n$-AlO(OH), $F(Q) = Q[S(Q) - 1]$ from Mo K$\alpha_1$ data (Fig. 1b); (b) PDF $G_{exp}(r)$ from (a); $Q_{max} = 130$ nm$^{-1}$, $r_{max} = 10$ nm, inset: $r_{max} = 4$ nm

**Local Structure Analysis**

Qualitatively, it easily can be seen from Fig. 3b that the PDF falls off to ~ 10% of its initial peak height around $r \approx 4$ nm. Since the instrumental resolution is of $0.02°/2\theta$, this fall-off is significantly due to the measured material. (For a comparison refer to the Ho-Mg-Zn alloys measured on the same Guinier instrument [7]: The FWHM is about 5 times smaller.) We can assume that beyond correlation lengths $r > 4$ nm the AlO(OH) structure begins to loose coherence. The crystallite size is thus directly reflected in the plotted PDF.

For quantitative analysis, atomic structure models can be built and a model PDF can be calculated from (4). $b_i$ is the scattering power of atom i, $\langle b \rangle$ is the average scattering power of the sample.

$$G(r)_{calc} \;=\; 1/r \sum_i \sum_j \left[\, b_i b_j / \langle b \rangle^2 \; \delta(r - r_{ij}) \,\right] - 4\pi r\, \rho_0 \qquad (4)$$

In a least squares refinement procedure using the software PDFFIT [8] the model can be verified like a periodic crystal structure in a Rietveld [9] refinement. The quantity to be minimized is a reliability factor

$$R = \sqrt{\{ \Sigma_{i=1}^{N} [G_{obs}(r_i)-G_{calc}(r_i)]^2 / \Sigma_{i=1}^{N} G^2_{obs}(r_i) \}} \qquad (5)$$

Compared to *R*-values obtained from Rietveld (or single crystal) refinements, *R*-values as defined in (5) for PDF refinements around 10 % are excellent. Good refinements with $R = 20$ to 30 % are reported [3]. When refining the PDF data, a certain *r*-range has to be specified since the atomic structure may differ on a local, medium or long-range scale. So local structures with different correlation lengths ξ are accounted for. ξ denotes the diameter of regions in which an identical atomic structure is present. These regions can be found all over the material but are not linked periodically. Thus the quality of the corresponding fit is dependent on *r* and reflected in $R = f(r_{min}, r_{max})$. *R*-values are comparable on the same data set but for different ranges and/or models.

To model *n*-AlO(OH) using the measured PDF, the structure data of crystalline Boehmite [6] were taken as a starting point: In its metric given above, three symmetry independent sites for non-hydrogen atoms in space group *Cmcm* are occupied, all on Wyckhoff position 4c 0 *y* ¼. Al(1) is located at $y = 0.683$; O(1) at $y = 0.286$ and O(2) at $y = 0.080$. O(2) is carrying the hydrogen atom which is not determined.

We chose three ranges of $\Delta r = 0.8$ nm to perform the refinements: **A** ($0.1 < r < 0.9$ nm), **B** ($0.9 < r < 1.7$ nm) and **C** ($1.7 < r < 2.5$ nm). First the scale factor and a parameter accounting for PDF peak widths, δ, were allowed to refine, then the lattice parameters *a*, *b* and *c*, an isotropic temperature factor $U_{eq}$ for each symmetry independent atom and finally the *y* coordinates of the atoms were refined. Figure 4 shows the respective measured, calculated and difference PDFs. Locally, the *Cmcm* model might be distorted: Therefore, symmetry reduction was applied to allow the atoms to deviate from $x = 0$ and $z = ¼$ (resulting in the *translationengleiche* subgroup *C*1), but this did not improve the fits. Table 2 comprises the results for the refinements in *Cmcm*. As the values for the **A** and **B** region do not differ significantly, the regions were combined (**AB**) and refined again. To account for the average structure, the combination **ABC** was refined as well.

Table 2: Least squares refinement results of *n*-AlO(OH) using PDFfit [8]

| range | **A** | **B** | **AB** | **C** | **ABC** | [6] |
|---|---|---|---|---|---|---|
| $r_{min} - r_{max}$/nm | 0.1 – 0.9 | 0.9 – 1.7 | 0.1 – 1.7 | 1.7 – 2.5 | 0.1 – 2.5 | (0 – ∞) |
| *R*/% | 29.6 | 33.8 | 28.3 | 44.5 | 44.7 | – |
| scale | 1.321(21) | 1.119(11) | 1.292(17) | 0.710(18) | 1.12(1) | – |
| δ | 0.808(7) | 2.1(3) | 0.818(7) | 6(2) | 0.82(3) | – |
| *a*/nm | 0.2879(3) | 0.28734(10) | 0.28754(17) | 0.28719(22) | 0.28758(13) | 0.28679(3) |
| *b*/nm | 1.2287(18) | 1.2291(7) | 1.2317(11) | 1.2271(8) | 1.2272(6) | 1.2214(1) |
| *c*/nm | 0.3714(5) | 0.37069(12) | 0.37006(25) | 0.3715(3) | 0.37146(15) | 0.36936(3) |
| *y*(Al(1)) | 0.6758(4) | 0.6747(9) | 0.6758(3) | 0.685(3) | 0.6738(6) | 0.683(1) |
| *y*(O(1)) | 0.3022(16) | 0.2999(14) | 0.3020(14) | 0.294(3) | 0.3000(10) | 0.286(2) |
| *y*(O(2)) | 0.0866(4) | 0.0861(8) | 0.0864(3) | 0.0742(22) | 0.0857(7) | 0.080(2) |
| $U_{eq}$(Al)/$10^{-4}$nm$^2$ | 0.038(6) | 0.0397(5) | 0.0386(5) | 0.0422(13) | 0.0415(6) | 0.12(4) |
| $U_{eq}$(O)/$10^{-4}$nm$^2$ | 0.0312(4) | 0.0303(4) | 0.0309(4) | 0.0263(8) | 0.0328(4) | 0.16(7) 0.13(6) |

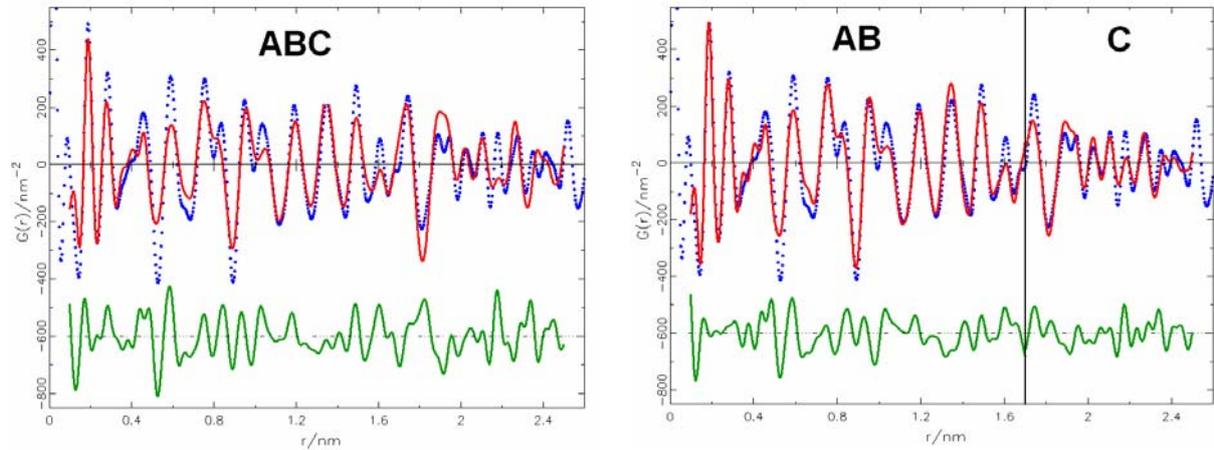

Figure 4: $G_{exp}(r)$ – blue dots, $G_{calc}(r)$ – red line, and their difference plots – green line, for the refined regions **ABC** ($R = 44.7\%$), **AB** ($R = 31.5\%$) and **C** ($R = 44.5\%$), respectively

**Discussion**

In the Boehmite structure, Al is surrounded by six O-Atoms in a distorted $AlO_6$ octahedron. Only four symmetry independent chemical bonds are present: 2 types Al(1)-O(1) (labelled a and b), one Al(1)-O(2) (labelled c) and one O(2)-O(2), the latter is the length of a hydrogen bridge and labelled d (Fig. 5a). Boehmite can be described as a layer structure: The basic layer is composed of two sheets of edge sharing $AlO_6$-octahedra, the sheets are fused again *via* common edges of the octahedra. Thus the aluminium atom is coordinated by four O(1) and two O(2) atoms, the O(1) atoms are inside the layer and are shared by adjacent $AlO_6$-octahedra. O(2) is connected to the hydrogen atoms, accommodated in between the layers, the layers are stacked along [010]. The layer structure and its constructive sheets are depicted in Fig. 5b, the distance of the layers is ½ b. More precisely, we define four 'sheet distances', $d_1$ to $d_4$ which are defined in Fig. 5b. It is $d_1 = [2y(O(1))–0.5]b$, $d_2 = [1.0–y(Al(1))–y(O(1))]b$, $d_3 = [y(Al(1))–y(O(2))–0.5]b$ and $d_4 = [2y(O(2))]b$; $y$ values have to be taken from Tab. 2.

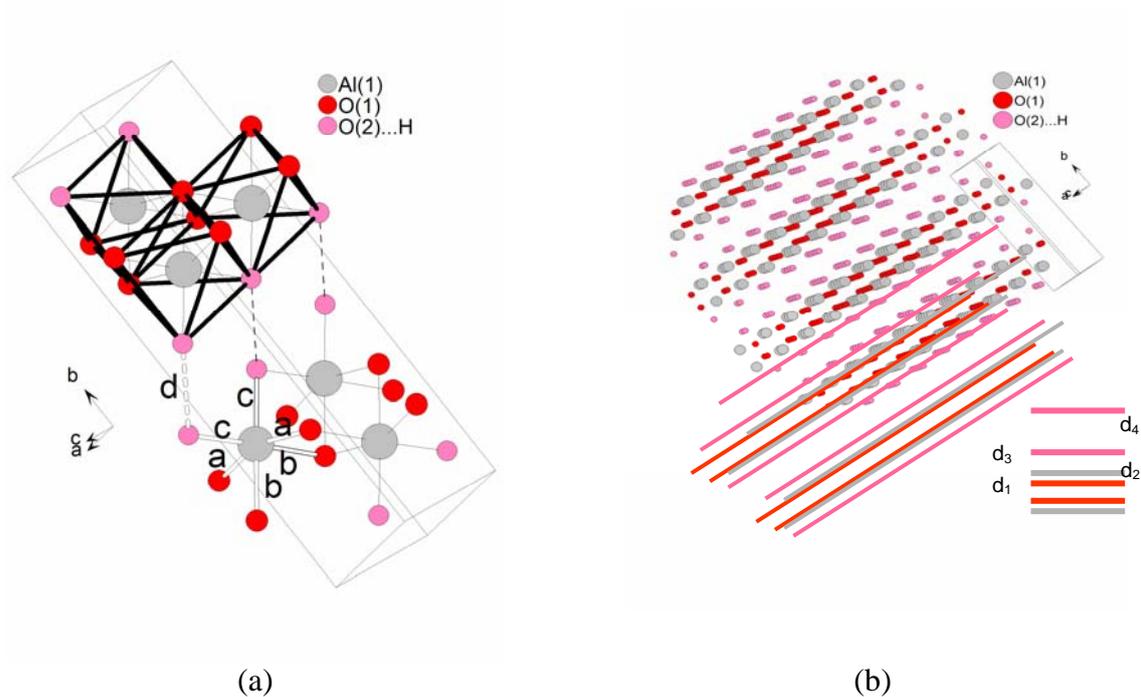

(a)            (b)

Figure 5: The Boehmite structure, view of a unit cell approximately along [101]: (a) arrangement of $AlO_6$-octahedra; four types of bond lenghts a, b, c and d are present in the structure; (b) a larger section visualizes the layer type structure, sheets separated by $d_1$ to $d_4$ are defined.

Table 3: Bond lengths and 'sheet distances' in pm for AlO(OH) derived from $r$-dependent PDF-refinements. The standard deviation ($\sigma$) given in column 'AB' is calculated from the refinement ranges **A**, **B** and **AB**.

| range | **A** | **B** | **AB** | **C** | **ABC** | [6] |
|---|---|---|---|---|---|---|
| $r_{min} - r_{max}$/nm | 0.1 – 0.9 | 0.9 – 1.7 | 0.1 – 1.7 | 1.7 – 2.5 | 0.1 – 2.5 | (0 – ∞) |
| a: Al(1)-O(1) | 187.7 | 188.0 | 187.0(1) | 187.5 | 188.5 | 188.5 |
| b: Al(1)-O(1) | 211.8 | 210.5 | 212(2) | 196.2 | 211.3 | 195.3 |
| c: Al(1)-O(2) | 180.9 | 180.3 | 181.1(3) | 197.8 | 179.9 | 190.8 |
| d: O(2)-H···O(2) | 282.4 | 281.3 | 282.0(6) | 260.1 | 280.6 | 268.9 |
| $d_1$ | 128 | 123 | 128(3) | 108 | 123 | 88 |
| $d_2$ | 27 | 31 | 27(2) | 26 | 32 | 38 |
| $d_3$ | 110 | 109 | 110(1) | 136 | 108 | 126 |
| $d_4$ | 213 | 212 | 213(1) | 182 | 210 | 195 |

Considering a crossover from the microcrystalline state of AlO(OH) adopting the Boehmite type structure (represented by [6]) to nanocrystallinity (represented by the range **AB**), one observes a distortion of the basic $AlO_6$ octahedron and an elongation of the interlayer hydrogen bridge about +5%. In Table 3 standard deviations $\sigma$ are calculated for the range AB from the refined values of the regions **A**, **B** and **AB**. The distortions observed are larger than $5\sigma$ and therefore are considered to be significant. This means an intrinsic change of the atomic and thus the layer structure inside the nano-particle. While bonds b and d enlarge, bond c diminishes and bond a keeps approximately constant (Tab. 3). This means both an increased separation of layers (the inter-layer distance $d_4$ increases about 10%) plus a more pronounced separation of the O-Al-O sheets (intra-layer distance $d_1$ increases about 40%). But, on the other hand, a contraction of the sheet O-Al-O ($d_2$ decreases by 40%) leaves the total stacking vector, identical to half of the *b* lattice parameter, virtually constant (less than 1%; *cf.* Tab. 2). Such a weakening of the hydrogen bond and break-up of the layers should account for the enhanced reactivity of *n*-AlO(OH).

The structural coherence in the investigated sample begins to cease about $\xi < 2$ nm as no longer all vectors of length $r > 2$ nm do not start and end inside the nano grain any more (Fig. 6). This is also evidently reflected in the large δ value for the **C** range which is faked by loss of coherence (δ is about 7.5 times larger compared to the **AB** range; see Tab. 2). The structure parameters, though, resemble more the values of the microcrystalline case [6] (Tab. 2). PDF-peaks for $r > 4.5$ nm are still present and point to preferred orientations of crystallites within the agglomerates.

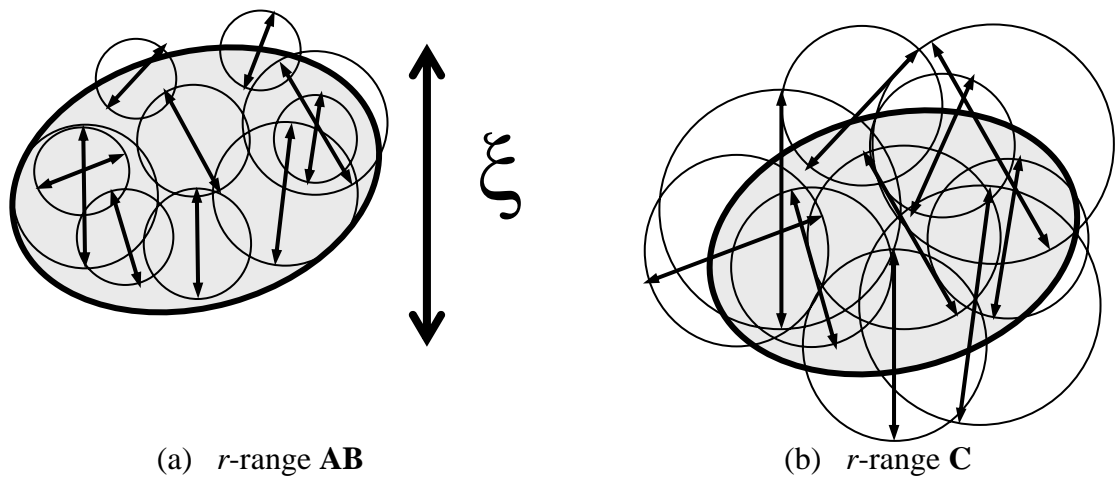

(a)    *r*-range **AB**                          (b)    *r*-range **C**

Figure 6: Schematic sketch of the maximal correlation ξ in a (hypothetically) oval particle. (a) 10 random vectors **r** representing the *r*-range **AB**, 10 vectors **r** in the *r*-range **C** at the similar relative positions w.r.t. the particle.

**Conclusions**

We present a straightforward approach to atomic structure determination of a nanomaterial using in-house X-ray powder diffraction data: The nanocrystallinity of Boehemite type Disperal® P2, *n*-AlO(OH), is clearly reflected in its PDF: Amplitudes of $G(r)$ fall off to less than 10% for $r > 4$ nm. Structural coherence is given for $r = 2$ nm $< \xi$ as follows from a sequence of least-squares local atomic structure refinements using $G(r)$ in different *r*-ranges. Inside the nano-particle, we observe contractions and distractions (–20 to +30 %) in the layer structure: This leads to a weakening of hydrogen bonding *inter*-layer by simultaneous retention of the layer stacking vector length, if compared to micro-crystalline material.
The results are in accordance with those obtained from a whole series of AlO(OH) of different crystallite sizes using Rietveld refinements [2].